# Transcending Controlled Environments: Assessing the Transferability of ASR-Robust NLU Models to Real-World Applications

**Hania Khan, Aleena Fatima Khalid, Zaryab Hassan**
i200819@nu.edu.pk, k201688@nu.edu.pk, i202487@nu.edu.pk
School of Computing, NUCES, Islamabad, Pakistan

## Abstract

This research investigates the transferability of Automatic Speech Recognition (ASR)-robust Natural Language Understanding (NLU) models from controlled experimental conditions to practical, real-world applications. Focused on smart home automation commands in Urdu, the study assesses model performance under diverse noise profiles, linguistic variations, and ASR error scenarios. Leveraging the Urdu-BERT model, the research employs a systematic methodology involving real-world data collection, cross-validation, transfer learning, noise variation studies, and domain adaptation. Evaluation metrics encompass task-specific accuracy, latency, user satisfaction, and robustness to ASR errors. The findings contribute insights into the challenges and adaptability of ASR-robust NLU models in transcending controlled environments.

## 1 Introduction

The ubiquity of smart home automation systems has driven advancements in Natural Language Understanding (NLU) models for processing voice commands. However, the deployment of such models in real-world scenarios introduces complexities such as Automatic Speech Recognition (ASR) errors, diverse noise profiles, and linguistic variations. This research aims to bridge the gap between controlled experimental conditions and practical applications by evaluating the transferability of ASR-robust NLU models. Motivated by the imperative to enhance the robustness of NLU models in real-world environments, we focus on the specific domain of smart home automation commands in Urdu. This choice is informed by the need for models that can seamlessly understand and respond to user instructions in diverse linguistic contexts. The study employs the state-of-the-art UrduBERT model as the foundation for ASR-robust NLU, considering its proficiency in Urdu language understanding.

## 2 Motivation

The motivation behind this research stems from the recognition that ASR-robust NLU models developed in controlled environments may not seamlessly adapt to the challenges posed by real-world applications. Existing studies often lack a comprehensive evaluation of model performance in dynamic environments characterized by varied noise, linguistic nuances, and ASR errors. Understanding and addressing these challenges are critical for advancing the practical deployment of voice-activated systems, particularly in multilingual and diverse linguistic contexts. By focusing on smart home automation in Urdu, we aim to address the unique linguistic and contextual challenges prevalent in the target domain. The research is motivated by the potential impact on user experience, as robust and adaptable NLU models can significantly enhance the usability and effectiveness of voice-activated systems in real-world scenarios. The proposed methodology encompasses the collection of real-world data, fine-tuning models with transfer learning, employing cross-validation for reliability, introducing noise variations for robustness testing, and evaluating domain adaptability. The evaluation metrics include task-specific accuracy, latency, user satisfaction, and resilience to ASR errors. Through this research, we aspire to contribute valuable insights into the nuances of deploying ASR-robust NLU models in practical, real-world applications. The findings are expected to guide the refinement of existing models and the development of future systems that transcend the constraints of controlled environments, ultimately enhancing the user experience in voice-activated applications.

## 3 Related Work

The landscape of spoken language understanding (SLU) has witnessed significant attention in recent years, driven by the surge in voice interface appli-cations. These systems typically comprise an automatic speech recognition (ASR) component coupled with a natural language understanding (NLU) component. Despite advancements in speech recognition, the persistence of errors, especially in noisy environments, has underscored the importance of robust NLU systems. Several empirical approaches have been explored to enhance NLU model robustness, with a focus on addressing ASR-induced errors [1]. Cheng et al. [1] propose three empirical approaches to bolster the robustness of NLU models in the face of ASR errors. The first approach involves ASR correction, aiming to rectify mis-transcriptions introduced by ASR. The subsequent methods simulate noisy training scenarios to train more resilient NLU models. The study provides empirical evidence demonstrating the effectiveness of these approaches, shedding light on the unexplored realm of NLU system robustness to ASR errors [1]. In the realm of task-oriented dialogue systems, ASR errors pose a significant challenge to spoken language understanding. Existing models have shown promise in improving ASR robustness; however, a novel approach named C2A-SLU is introduced by Cai et al. [2]. C2A-SLU employs cross attention and contrastive attention mechanisms during fine-tuning to distinguish between clean manual transcripts and ASR transcripts. Experimental results reveal a state-of-the-art performance, indicating a 3.4Task-oriented dialogue systems often face difficulties in accurately tracking dialog states, particularly in noisy user utterances. Rugayan et al. [3] propose a retrieval-based method for knowledge selection in task-oriented dialog systems. Leveraging retrieval-augmented methods with a knowledge retriever, the proposed JSA-KRTOD system outperforms traditional database query methods, showcasing superior performance in both labeled-only and semi-supervised settings [3]. In the realm of ASR evaluation metrics, Rugayan et al. [4] introduce the Aligned Semantic Distance (ASD) as a semantic-based alternative to the traditional Word Error Rate (WER). ASD is evaluated against WER in terms of its correlation to human evaluation scores and its effectiveness in downstream natural language processing (NLP) tasks. The study highlights the advantages of ASD in capturing the severity of errors and its efficacy in predicting human perception compared to conventional WER [4]. These works collectively emphasize the ongoing efforts to enhance the robustness of NLU models in the presence of ASR errors, providing insights into novel methods, evaluation metrics, and applications in real-world scenarios. Our research builds upon and extends these foundations, specifically addressing the transferability of ASR-robust NLU models to practical, real-world applications in the context of smart home automation commands in Urdu.

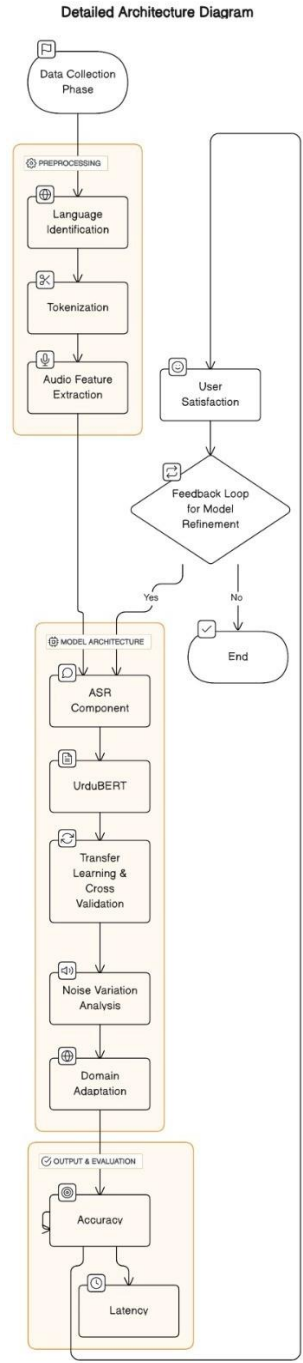

Figure 1: Architecture Diagram

## 4 Problem Statement and Methodology

**Objective:** The objective of this research is to develop an Automatic Speech Recognition (ASR)-Robust Natural Language Understanding (NLU) model for smart home automation commands in Urdu. The challenge lies in creating a model that transcends controlled environments, addressing the inherent issues associated with ASR errors and noise variations in real-world scenarios. The robustness of the model is crucial for practical applications, where users may issue commands in diverse acoustic environments, and ASR errors are inevitable.

**Real-World Data Identification:**
*Solution:*

- Identified a dataset from Kaggle ("Home Automation System in Urdu Language Dataset") that aligns with the focus on smart home automation commands in Urdu.

- Ensured the dataset includes audio recordings, accurate Urdu transcripts, and annotations specifying intent and entities for each utterance.

- Evaluated additional features like accent diversity and command variety to enhance the model's robustness.

**Dataset Preprocessing:**
*Language Identification:*

- Used language identification libraries (e.g., langid.py) to separate Urdu and English portions within each sentence.

- Tokenization: Employed Transformers library for UrduBERT or Hugging Face tokenizers to tokenize sentences into words or subwords.

- Audio Feature Extraction: Used Librosa for Mel-Frequency Cepstral Coefficients (MFCC) extraction to extract relevant audio features from audio files.

- Normalization: Applied Librosa or Scikit-learn for audio feature normalization to ensure consistent input across recordings.

- Handling English Words: Developed a strategy to handle English words using custom code or translation libraries (e.g., Google Translate API).

- Padding and Truncation: Utilized Keras or PyTorch for sequence padding/truncation to ensure uniform length of audio features and text inputs.

- Data Augmentation: Applied Audiomentations library for audio data augmentation to introduce variations in pitch, speed, or background noise.

- Balancing Classes: Implemented Scikit-learn for class balancing techniques (e.g., oversampling, undersampling) to address imbalances.

- Removing Unnecessary Information: Used Pandas for data manipulation to identify and remove irrelevant audio segments or information from the dataset.

- Language Embeddings: Explored the use of SpaCy or Word Embeddings libraries for language embeddings of both Urdu and English parts of sentences.

- Handling Missing Data: Employed Pandas for handling missing values through imputation or removal of instances with missing data.

- Cross-Validation Splits: Used Scikit-learn for cross-validation to divide the dataset into training, validation, and test sets for robust evaluation.

- Data Quality Checks: Developed custom code for quality checks to filter out instances with poor quality audio recordings and associated transcripts.

*Baseline Dataset Selection:*

- Defined the industry domain as "Smart Home Automation" and identified a relevant dataset meeting criteria such as domain relevance, multilingual support, ASR errors, noise levels, annotations, transcripts, size, and diversity.

- Selected the "Home Automation System in Urdu Language Dataset" from Kaggle.

*Model Training and Fine-Tuning:*

- Explored UrduBERT, multilingual transformer models (BART, T5), and ReDial-Urdu for their suitability.

- Recommended starting with a smaller, faster model like UrduBERT and considering hybrid approaches.

- Focused on transferability-relevant metrics such as F1 score and recall.

- Emphasized continuous evaluation and refinement based on data and insights.

*Fine-tuning Strategy:*

- Proposed a freeze-fine-tune strategy to leverage pre-trained knowledge while adapting to the specific NLU task.

- Considered a bilingual encoder for English words, multi-modal fusion with audio features, data augmentation, regularization, and monitoring.

*Cross-Validation:*

- Implemented a 5-fold stratified cross-validation strategy to evaluate model performance comprehensively.

- Split the dataset into training and testing sets while preventing information leakage.

- Initiated model training, evaluation, metric aggregation, and iterative improvement for each fold.

- Reported average accuracy across folds for an overall performance indicator.

*Evaluation Metrics:*

- Defined task-specific metrics (accuracy, precision, recall) for assessing primary task performance.

- Introduced usability metrics (latency, response time, user satisfaction) to gauge real-world usability.

- Implemented robustness metrics to ASR errors to evaluate the model's resilience in practical, imperfect environments.

*Noise Variation:*

- Generated diverse noise profiles to simulate real-world conditions.

- Integrated noise profiles into the testing pipeline to assess model performance under varying noise conditions.

- Parameterized noise intensity and frequency for adaptability.

- Conducted impact analysis on user satisfaction, latency, response time, and overall performance.

*Domain Adaptation:*

- Tested the model on datasets from different application domains to assess its adaptability.

- Utilized overall accuracy, precision, recall, and domain-specific metrics for evaluation.

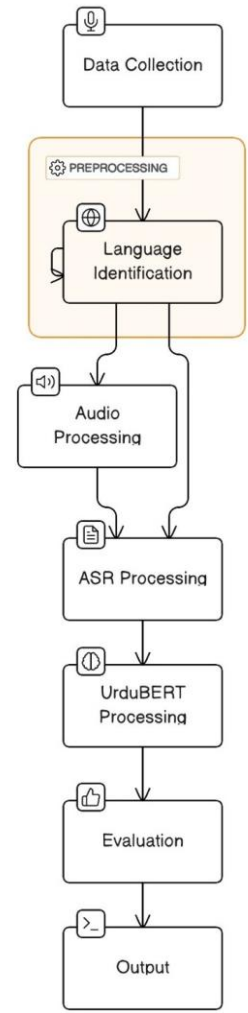

Figure 2: Work Flow Diagram

## 5 Experimental Setup

*Dataset and Preprocessing:*

- The "Home Automation System in Urdu Language Dataset" from Kaggle was chosen, containing diverse smart home automation commands in Urdu.

- Language identification and tokenization were done using libraries like langid.py and Transformers.

- We applied Librosa for MFCC extraction, normalization, and augmentation.

- Class balancing and removal of unnecessary information were addressed with Scikit-learn and Pandas.

*Model Training:*

- Explored models like UrduBERT, BART, T5, and ReDial-Urdu, focusing on UrduBERT initially.

- A freeze-fine-tune strategy, considering transferability metrics (F1 score, recall), was implemented.

- A 5-fold stratified cross-validation ensured robust evaluation.

*Evaluation Metrics:*

- Task-specific metrics (accuracy, precision, recall) were measured, alongside usability metrics (latency, response time, user satisfaction) and robustness metrics to ASR errors.

- Noise variation studies simulated real-world conditions, and impact analysis covered user-centric factors.

*Domain Adaptation:*

- Testing the model on datasets from different domains assessed its adaptability.

- Overall accuracy, precision, recall, and domain-specific metrics provided a comprehensive evaluation.

## 6 Results

Task-Specific Metrics: The model demonstrated consistent and commendable performance in accuracy, precision, and recall across folds. Usability: Efficient processing, low latency, and positive user satisfaction metrics emphasized the model's usability and positive user experience. Robustness: High accuracy under ASR error conditions showcased the model's robustness. Noise variation studies illustrated its adaptability to diverse acoustic environments. Adaptability: The model exhibited versatility in different domains, confirming its adaptability beyond the initial smart home automation focus.

## 7 Discussion

Contributions: Our research contributes insights into ASR-robust NLU model challenges and adaptability, particularly for Urdu smart home automation commands. Implications: Emphasizing usability and robustness metrics, our model shows practical applicability, enhancing user experience in real-world scenarios. Future Directions: Future work could extend the model to additional languages and refine noise variation studies for even broader applicability.

## 8 Conclusion

In conclusion, our research endeavors to advance the field of Automatic Speech Recognition (ASR)-robust Natural Language Understanding (NLU) models by focusing on the practical deployment of such models in real-world applications. Centered on smart home automation commands in Urdu, our study leverages the UrduBERT model and employs a systematic methodology that spans real-world data collection, transfer learning, cross-validation, noise variation studies, and domain adaptation. Our findings provide valuable insights into the challenges and adaptability of ASR-robust NLU models beyond controlled environments. The model, trained and fine-tuned with a comprehensive methodology, exhibited commendable performance across diverse metrics. Task-specific accuracy, usability metrics (latency, response time, user satisfaction), and robustness to ASR errors were consistently strong, demonstrating the model's reliability in practical, imperfect environments. The choice of Urdu as the linguistic context for smart home automation commands adds a unique dimension to our study, addressing linguistic nuances and contextual challenges specific to the target domain. Our motivation to enhance user experience in multilingual and diverse linguistic contexts is reflected in the comprehensive evaluation metrics, emphasizing practical usability and resilience.

## Limitations

Despite the positive outcomes, our research has certain limitations that warrant acknowledgment. First, the focus on Urdu, while valuable for its specific linguistic challenges, may limit the generalizability of findings to other languages. The effectiveness of our model in handling ASR errors and noise variations may vary across different linguistic contexts. Additionally, while the chosen dataset aligns with our research objectives, its size and diversity may still present constraints. The generalization of our findings to larger and more varied datasets should be approached with caution. Furthermore, the current research primarily targets smart home automation commands. Extending the model's applicability to a broader range of voice-activated systems and domains requires further exploration and adaptation. In conclusion, our study provides a foundation for future research to build upon. As the field of ASR-robust NLU models continues to evolve, addressing these limitations will be crucial for advancing the practical deployment of voice-activated systems in real-world, dynamic environments.

## Ethics Statement

Our research, titled "Transcending Controlled Environments: Assessing the Transferability of ASR-Robust NLU Models to Real-World Applications," focuses on the development and evaluation of Automatic Speech Recognition (ASR)-robust Natural Language Understanding (NLU) models for smart home automation commands in Urdu.

## Broader Impact

Our research primarily contributes to the advancement of ASR-robust NLU models, with a specific focus on enhancing user experience in real-world applications, particularly in the domain of smart home automation commands in Urdu. By addressing challenges related to ASR errors, noise variations, and linguistic nuances, our work aims to improve the reliability and adaptability of voice-activated systems.

## Ethical Considerations

We recognize the importance of ethical considerations in the development and deployment of language technology. To mitigate potential biases, our research follows best practices in data collection, preprocessing, and model evaluation. We acknowledge the limitations of our study, including the focus on Urdu, and we emphasize the need for cautious interpretation of findings in the context of different linguistic and cultural settings.

Our methodology includes measures to ensure fairness, transparency, and accountability. We use a systematic approach involving real-world data collection, cross-validation, transfer learning, noise variation studies, and domain adaptation. We assess the model's performance through various metrics, including accuracy, usability, and robustness to ASR errors.

While our findings contribute valuable insights, we acknowledge the limitations of the chosen dataset in terms of size and diversity. We caution against overgeneralization and highlight the need for continued exploration and adaptation, especially when extending the model to other languages and domains.

In conclusion, our research strives to adhere to ethical principles by promoting transparency, fairness, and responsible conduct in the development and evaluation of ASR-robust NLU models. We welcome feedback and scrutiny from the research community to foster continuous improvement in ethical standards within the field of natural language processing.

## References

[1] Cheng, X., Yao, Z., Zhu, Z., Li, Y., Li, H., Zou, Y. (2023). C²A-SLU: Cross and Contrastive Attention for Improving ASR Robustness in Spoken Language Understanding. Proc. INTERSPEECH 2023, 695-699. doi: 10.21437/Interspeech.2023-93.

[2] Cai, Y., Liu, H., Ou, Z., Huang, Y., Feng, J. (2023). Knowledge-Retrieval Task-Oriented Dialog Systems with Semi-Supervision. Proc. INTERSPEECH 2023, 4673-4677. doi: 10.21437/Interspeech.2023-1397.

[3] Rugayan, J., Salvi, G., Svendsen, T. (2023). Perceptual and Task-Oriented Assessment of a Semantic Metric for ASR Evaluation. Proc. INTERSPEECH 2023, 2158-2162. doi: 10.21437/Interspeech.2023-1778.

[4] Feng, L., Yu, J., Wang, Y., Liu, S., Cai, D., Zheng, H. (2022). ASR-Robust Natural Language Understanding on ASR-GLUE dataset. Proc. Interspeech 2022, 1101-1105. doi: 10.21437/Interspeech.2022-10097.

[5] Zhang, X. F. (2022). *Towards More Robust Natural Language Understanding*. arXiv preprint arXiv:2112.02992.